\documentclass[12pt]{article}
\usepackage[pctex32]{graphics}
\begin{document}
\vspace{1cm}
\begin{center}
~
\\
~
{\bf  \Large Comment on ``Quantum Statistical Mechanics of an Ideal Gas with Fractional Exclusion Statistics in Arbitrary Dimension"}
\vspace{1cm}

                      Wung-Hong Huang\\
                       Department of Physics\\
                       National Cheng Kung University\\
                       Tainan, Taiwan\\

\end{center}
\vspace{2cm}
It is mentioned that anyon thermodynamic potential $Q(\alpha, N)$ could not be factorized in terms characteristic of the ideal boson $\alpha =0$ and fermion $\alpha =1$ gases by the relation $Q(\alpha, N) = (1-\alpha) Q(0, N_b)+ \alpha Q(1, N_f)$ in which $N=N_f +N_b$, that claimed in Phys. Rev. Lett. 78, 3233 (1997).  Our analyses indicate that the thermodynamic quantities of anyon gas may be factorized as $Q(\alpha) = \alpha Q(1) + (1-\alpha) Q(0)$ only in the two-dimension system. 
\vspace{2cm}
\begin{flushleft}
Phys. Rev. Lett.  81(1998)2392
\\
*E-mail:  whhwung@mail.ncku.edu.tw\\
\end{flushleft}
\newpage
In a recent Letter on quantum statistical mechanics of an ideal gas with fractional exclusion statistics, Iguchi [1] proved an interesting property that anyons can be 
regarded as composites of fermions and boson in arbitrary dimension, i.e., in an N-anyon system the thermodynamic potential $Q(\alpha, N)$ can be factorized in terms characteristic of the ideal boson $\alpha =0$ and fermion $\alpha =1$ gases by the relation
$$Q(\alpha, N) = (1-\alpha) Q(0, N_b)+ \alpha Q(1, N_f),$$
in which $N=N_f +N_b$. In this Comment I first use high-temperature perturbation theory to point out that this relation does not exit in an arbitrary dimension. Next, I prove that in the two-dimensional N-anyon system the 
thermodynamic potential can be factorized as 
$$Q(\alpha, N) = (1-\alpha) Q(0, N)+ \alpha Q(1, N)$$.

First, from the anyon distribution function $n=1/(W +\alpha)$, and the relation [2]
 $$W^\alpha (1+W)^{1-\alpha} = e^{\beta(\epsilon-\mu)},$$
the high-temperature expansion of the energy becomes 
$$E(a,N)={D\over2}NkT\left[1+{N\over V}\lambda^D (-1+2\alpha)/s^{1+D/2}\right.$$
$$\left.+ \left({N\over V}\lambda^D\right)^2[1/2^D -2/3^{(1+D/2)}-4\alpha/2^D+3\alpha/3^{D/2)}+4\alpha^2/2^D-3\alpha^2/3^{D/2}]+....\right]$$
in which $\lambda$ is the conventional thermal wavelength, and $D$ is the space dimension. From this result we see  that the relation $Q(\alpha, N) = (1-\alpha) Q(0, N_b)+ \alpha Q(1, N_f),$ cannot be satisfied for any value of $(N_f, N_b)$ with the condition $N=N_f +N_b$. Thus the theorem 3 in [1] is incorrect. The mistake can be traced to the fact that the definitions $W(\epsilon) = e^{\beta(\epsilon-\mu_f)}$ and $1+W(\epsilon) = e^{\beta(\epsilon-\mu_b)}$  
used in [1] can be satisfied only for  $\epsilon=0$ but not for other values of $\epsilon$.

Next, I have evaluated the high-temperature expansion of the energy $E(\alpha , N)$ to order  $(\lambda^D)^7$, and the result shows that only in two dimensions can it be linear in $\alpha$. Let me now prove that energy $E(\alpha , N)$ is linear in $\alpha$. Using the definition of $W$  we have two useful relations: 
$$d[ln(1+W^{-1})]/d\epsilon = -\beta n$$
$$W^{-1} (1+W)^{-1} dW/d\alpha = n [-\beta d\mu/d\alpha + ln(1+W^{-1})]$$
Then, as the density of state $N_d$ is constant in two dimensions we have the relation 
$$d[PV\beta]/d\alpha=\int d\epsilon N_d d[ln(1+W^{-1})0/d\alpha = -\int d\epsilon N_d W^{-1} (1+W)^{-1} dW/\alpha$$
$$ = -\int d\epsilon N_d n [-\beta d\mu/\alpha +ln(1+W^{-1})]$$
$$=N \beta d\mu/d\alpha -kT\int N_d ln(1+W^{-1}d[ln(1+W^{-1})]/d\epsilon$$
$$=N\beta d\mu/d\alpha +{kT\over2} N_d [ln(1+W(0)^{-1})]^2$$
Now, because $\mu$ is linear in $\alpha$ [2] and $W(0) = e^{-\beta\mu_b}$, 
we thus see that $d[PV\beta]/d\alpha$ does not depend on $\alpha$. This means that the thermodynamic potential $Q(\alpha, N)$ is linear in a and can be factorized as $Q(\alpha, N) = (1-\alpha) Q(0, N)+ \alpha Q(1, N)$.

In summary, the relation $$Q(\alpha, N) = (1-\alpha) Q(0, N_b)+ \alpha Q(1, N_f),$$ where $N=N_f +N_b$ , does not hold in an arbitrary dimension anyon system, while in a two-dimensional anyon system the thermodynamic 
potential can be factorized as $Q(\alpha, N) = (1-\alpha) Q(0, N)+ \alpha Q(1, N)$.
\\
~
\\
(Remarks: The factorizable property could be explained from the equivalence 
between the anyon statistics and statistics in a system with boson-fermion transmutation [3], as that discussed in [4].)
\\
~
\\
~
\\
~
{\bf  \Large References}
\begin{enumerate}
\item K. Iguchi, ``Quantum Statistical Mechanics of an Ideal Gas with Fractional Exclusion Statistics in Arbitrary Dimensions," Phys. Rev. Lett. 78, 3233 (1997). 
\item Y. S. Wu, ``Statistical Distribution for Generalized Ideal Gas of Fractional-Statistics Particles," Phys. Rev. Lett. 73, 922 (1994).
\item Wung-Hong Huang, ``Boson-Fermion Transmutation and Statistics of Anyon," Phys. Rev. E51 (1995) 3729  [hep-th/0308095].
\item Wung-Hong Huang, ``Statistics of Anyon Gas and the Factorizable Property of Thermodynamic Quantities,"  Phys. Rev B53 (1996) 15842 [hep-th/0702070].

\end{enumerate}
\end{document}